\documentclass[a4paper, journal, onecolumn]{IEEEtran}
\usepackage[dvips]{graphicx}
\usepackage{ifpdf}
\usepackage[cmex10]{amsmath}
\usepackage{amssymb,dsfont,mathrsfs}
\usepackage[tight,footnotesize]{subfigure}
\usepackage{multirow}
\usepackage{color}
\usepackage{algorithm}
\usepackage{algpseudocode}
\usepackage{hyperref}
\floatname{algorithm}{Algorithm}

\usepackage{mathtools}

\ifCLASSINFOpdf
\else
\fi

\begin{document}
%
\title{GuardRider: Towards Sustainable Backscattering System over WiFi in the Wild}

\author{
Xin~He, Weiwei~Jiang, Meng~Cheng, Xiaobo~Zhou, Peng-Jun~Wan, Panlong~Yang and Brian~Kurkoski

\thanks{X. He is with the School of Computer Science and Technology, University of Science and Technology, Hefei, China and is also with the School of Computer and Information, Anhui Normal University, Wuhu, China. (email:xin.he@ahnu.edu.cn).}
\thanks{W. Jiang is with the School of Computing and Information Systems, University of Melbourne. (email: weiwei.jiang@student.unimelb.edu.au).}
\thanks{X. Zhou is with the School of Computer Science and Technology, University of Tianjin, Tianjin, China. (email: xiaobo.zhou@tju.edu.cn).}
\thanks{M. Cheng and B. Kurkoski are with the School of Information Science, Japan Advanced Institute of Science and Technology, Ishikawa, Japan. (email: \{m-cheng, kurkoski\}@jaist.ac.jp).}
\thanks{P. Yang is with the School of Computer Science and Technology, University of Science and Technology, Hefei, China. (email: plyang@ustc.edu.cn).}
\thanks{P. Wan is with the Department of Computer Science, Illinois Institute of Technology, Chicago, IL 60616 USA (email: wan@cs.iit.edu).}
}

\maketitle

\begin{abstract}
The WiFi backscatter communications offer ultra-low power and ubiquitous connections for IoT systems. Caused by the intermittent-nature of the WiFi traffics, state-of-the-art WiFi backscatter communications are not reliable for backscatter link or simple for tag to do adaptive transmission. In order to build sustainable (reliable and simple) WiFi backscatter communications, we present GuardRider, a WiFi backscatter system that enables backscatter communications riding on WiFi signals in the wild. The key contribution of GuardRider is an optimization algorithm of designing RS codes to follow the statistical knowledge of WiFi traffics and adjust backscatter transmission. With GuardRider, the reliable baskscatter link is guaranteed and a backscatter tag is able to adaptively transmit information without heavily listening the excitation channel. We built a hardware prototype of GuardRider using a customized tag with FPGA implementation. Both the simulations and field experiments verify that GuardRider could achieve a notably gains in bit error rate and frame error rate, which are hundredfold reduction in simulations and around $99\%$ in filed experiments. 
\end{abstract}


\IEEEpeerreviewmaketitle
\hfill\par


\section{Introduction}
Backscatter communication is recognized as a promising technology for connecting the Internet of things (IoT) devices, due to its energy efficiency. A fundamental backscattering communication system consists of three components: an excitation source, a backscatter tag, and a legacy receiver. Excited by the radio frequency (RF) signals\footnote{Hereafter, these signals are referred to as excitation signals in backscatter systems.} emitted from the excitation source, the backscatter tag can transmit data to the receiver by modulating the excitation signals through changing its load impedance. Concisely, the backscatter tag does not generate carrier signals itself, but borrows carrier signals from the excitation source. It thus consumes only mircowatts of power during the backscatter stage \cite{FreeRider}.


Ambient backscatter leverages the RF signals from surrounding RF transmitters, which does not require a dedicated infrastructure as in the traditional backscatter systems \cite{AmBack13}. The backscatter tag could reflect TV signals \cite{AmBack13}, FM radio \cite{Wang17}, Bluetooth \cite{Iyer16},  WiFi signals \cite{Kellogg16}, etc., which are already around us. Among them, WiFi backscatter communication is gaining momentum because WiFi signals are ubiquitous \cite{BackFi, Bryce14, HitchHike, Kim18}. The WiFi backscatter communication can be viewed as a ``rider'' system since the information of the tag rides on the excitation signals. The advantages of rider mode are:
\begin{itemize}
\item {\it Low deployment cost.} It can ease the deployment of ambient backscatter communication systems without additional infrastructure.
\item {\it Ubiquitous deployment.} With wide deployment of access points (APs), WiFi signals are available in most residential areas, commercial zones, transportation stations, etc.
\end{itemize}

\begin{figure}
  \centering
  \includegraphics[width=3.25in]{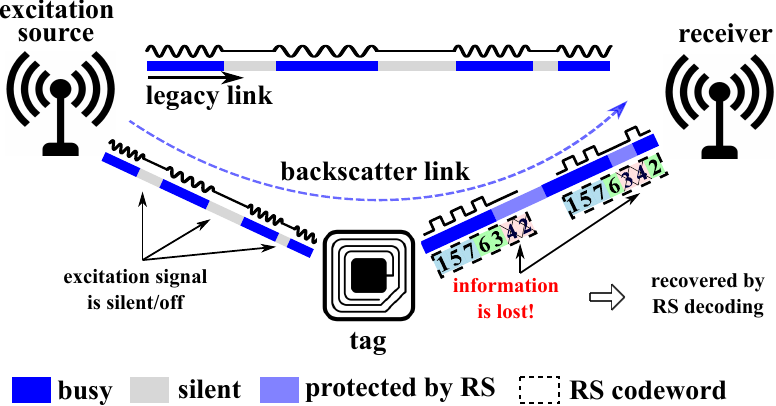}
  \caption{GuardRider operates using RS code to enable backscattering over the WiFi signal in the wild. Assume that the information is encoded to a codeword $\{1,5,7,6,3,4,2\}$ for backscattering.}
  \label{fig:sys}
\end{figure}

However, it has been found that WiFi traffics inherently have silent periods with varying lengths \cite{Talla15, Pareto} because of the following two factors. Firstly, a short interframe space (SIFS) is appended after each frame based on 802.11 protocols. The time of SIFS is typically at mircosecond level, e.g., the SIFS for 802.11a is 16$\mu$s \cite{80211}. Secondly, the traffic load of the wireless networks and its adopted higher layer protocols further make the durations of silent period constantly changing \cite{Sinha}. Moreover, the durations of the silent state are not identically distributed over time. Caused by these factors, the backscatter communication is {\it unreliable} since it suffers from high frame drop rate. Besides, it is extremely difficult for the backscatter tag to track the change of the silent states and adjust the backscatter transmission due to the low-power constraint of the tag. Therefore, in order to rider over WiFi signals in the wild, the on and off states of WiFi signals do need to be fully taken into account.

To this end, we present GuardRider, a WiFi backscatter system enables {\it sustainable backscatter transmission} which rides over WiFi signals using the optimized Reed-Solomon (RS) codes, as shown in Fig.~\ref{fig:sys}. With GuardRider, the backscatter tag is able to transmit information with relatively high reliability over ubiquitous WiFi signals. No negotiation is needed between the tag and the excitation source, which results in a fully rider system. The key insight is that the backscatter link can be treated as a burst error channel since the erased bits are consecutive caused by the silent period. Inspired by burst error correcting codes, we use adaptive RS codes to fit the mean time span of the silent period for enabling reliable backscatter transmission. To design and implement GuardRider, we face the following challenges:
\begin{itemize}
  \item {\bf Predict the capricious pattern of silent states.} It is extremely difficult to predict the pattern of the WiFi signals, especially for a low-power backscatter tag. The tag can only backscatter the information blindly without the knowledge of WiFi signals, or the WiFi transceiver has to send specific controlling signals to the network. Blind backscatter leads to poor system performance, while the latter one has interruption to the legacy link. Moreover, design of RS codes highly depends on the pattern of WiFi signals, and then predicting the pattern of the WiFi signal is the first important step.
  \item {\bf Strike a balance between efficiency and reliability of the backscatter link.} In practice, the silent period of WiFi signals varies significantly, and the number of redundant symbols should change accordingly. The unchanging RS code leads to low efficiency or low reliability. More concretely, if redundant symbols in RS code are inserted excessively, high reliability is guaranteed, while the bandwidth is wasted\footnote{The bandwidth is a scarce resource of backscatter communications.}. Conversely, if the redundant symbols are insufficient, the backscatter system is of low reliability. To meet the balance between the efficiency and reliability, the tag should be adaptive to the excitation channel with low complexity.
\end{itemize}

{\it Summary of results and contributions.} Our system addresses the above challenges by modeling the distribution for the silent periods of WiFi signals using maximum likelihood estimation (MLE), and optimizing the code design of RS codes appropriately based on a heuristic algorithm. We built a hardware prototype using a customized tag and implemented our system with a universal software radio peripheral (USRP) reconfigurable I/O (RIO) device with a field-programmable gate array (FPGA). The obtained results from conducted simulations and experiments verify that GuardRider could achieve the following outcomes.
\begin{itemize}
  \item {\bf System level-simplicity.} GuardRider is the first proposed backscatter communication system that carefully examines the characteristics on the alternation of on-and-off states of the excitation WiFi signals, and designs an adaptive RS code to follow the characteristics. GuardRider shifts the estimating process of the on-and-off states of WiFi signals from the tag to the receiver, and thus the backscatter tag could be operated without increasing any additional cost for predicting WiFi traffics.
  \item {\bf Link level-reliability.} From the results of simulations, the bit error rate (BER) performance of GuardRider is reduced by hundredfold (e.g., reduced to $10^{-6}$ from $10^{-4}$). Moreover, in the experiments, the FER performances of GuardRider are also reduced significantly compared to the baseline system (e.g., reduced to 0.25 from 1). In  general, GuardRider significantly improves the reliability of backscatter communications, which could be considered as a potential solution to the IoT applications requiring high reliability.
\end{itemize}

The rest of the paper is organized as follows. Section II provides the preliminary knowledge about RS codes. The proposed optimization algorithm of designing RS codes is presented in Section III, followed by the system design and implementation of GuardRider in Section IV. The performance of GuardRider is evaluated in Section V. The discussion and related work are given in Section VI and VII, respectively. Finally, we conclude this paper in Section VIII with several concluding statements.

\section{Preliminaries}
\begin{figure}
  \centering
  \includegraphics[width=3in]{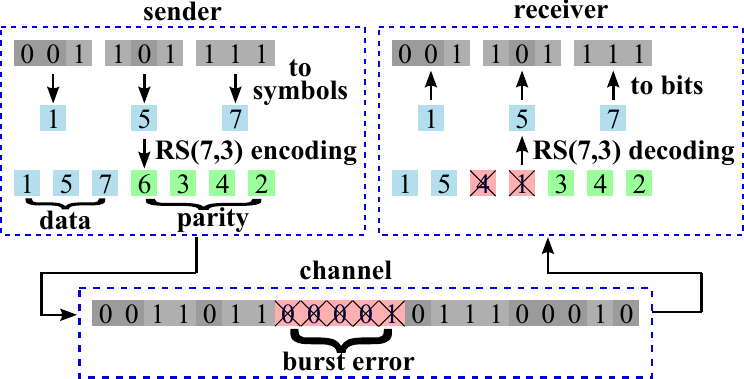}
  \caption{Example of RS encoding and decoding.}\label{fig:rs}
\end{figure}
\subsection{Reed-Solomon (RS) codes}
As mentioned above, the WiFi traffic is intermittent, which allows us to model the backscatter link as a burst error channel. Reed-Solomon (RS) codes are block-based error correcting codes with a wide range of applications in digital communications and storage, with the aim of correcting burst and/or erasure errors.

\begin{figure*}
  \centering
  \subfigure[packet size 128 bytes]{
        \includegraphics[width=0.22\textwidth]{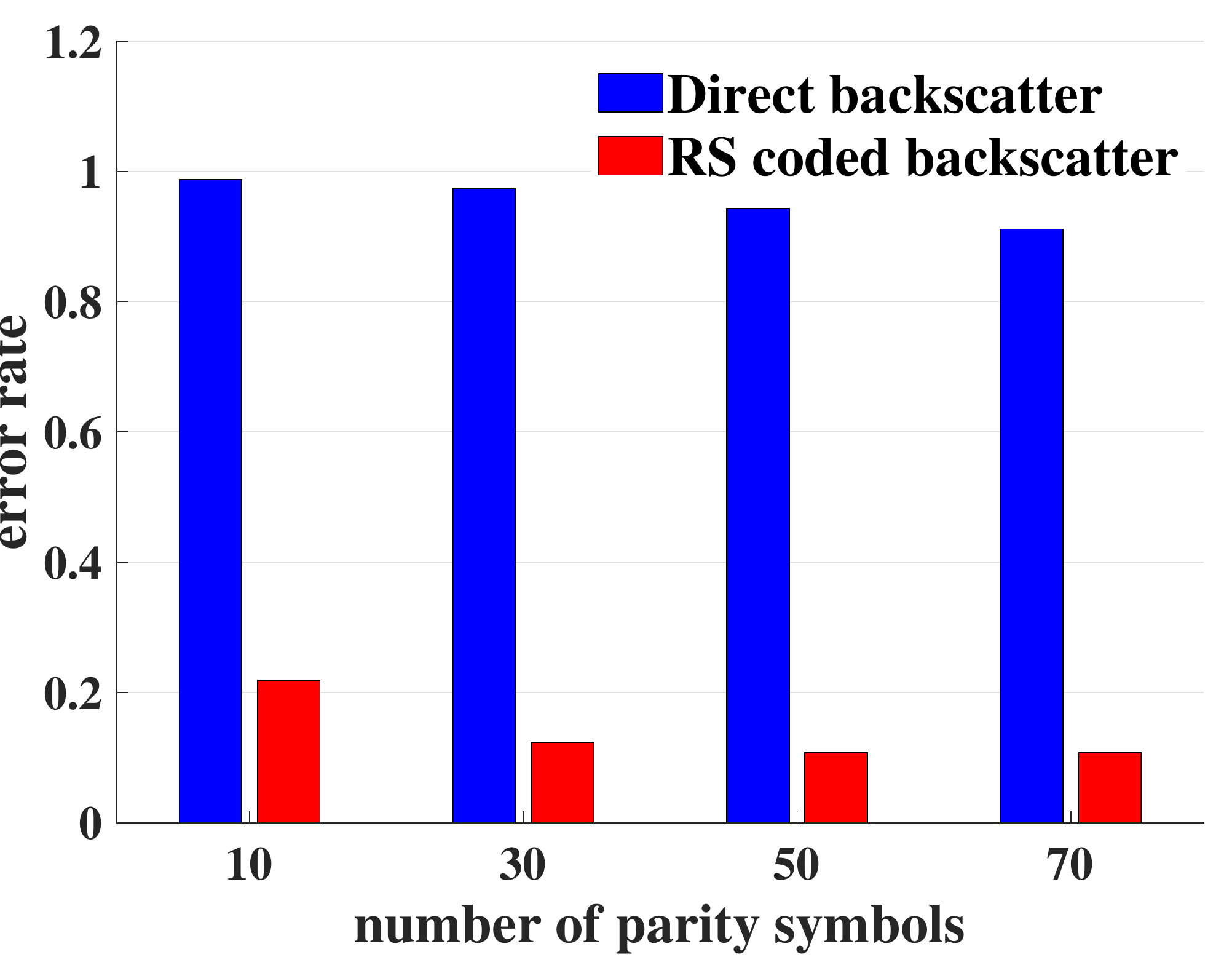}}
  ~
  \subfigure[packet szie 512 bytes]{
        \includegraphics[width=0.22\textwidth]{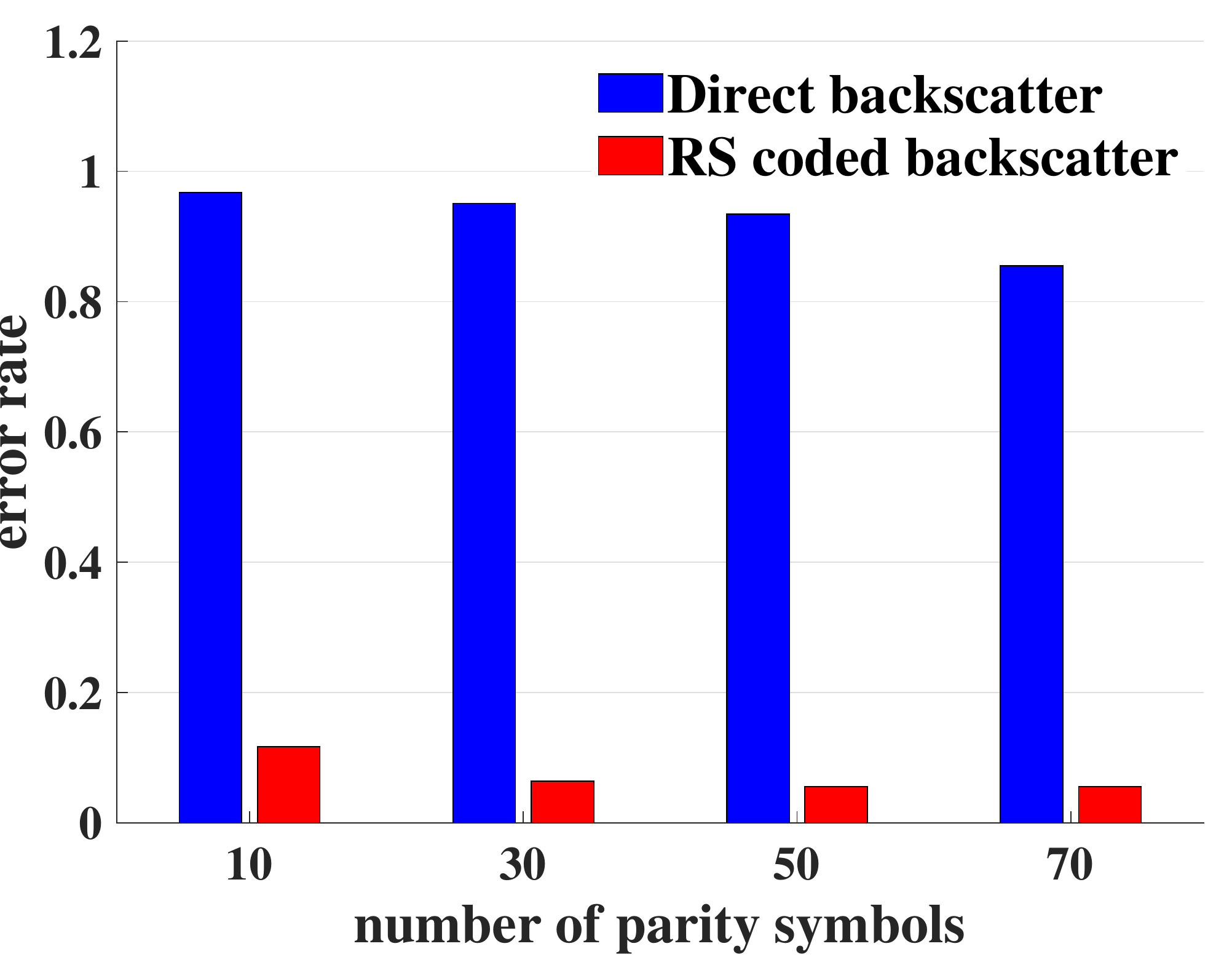}}
  ~
  \subfigure[WiFi traffic]{
        \includegraphics[width=0.22\textwidth]{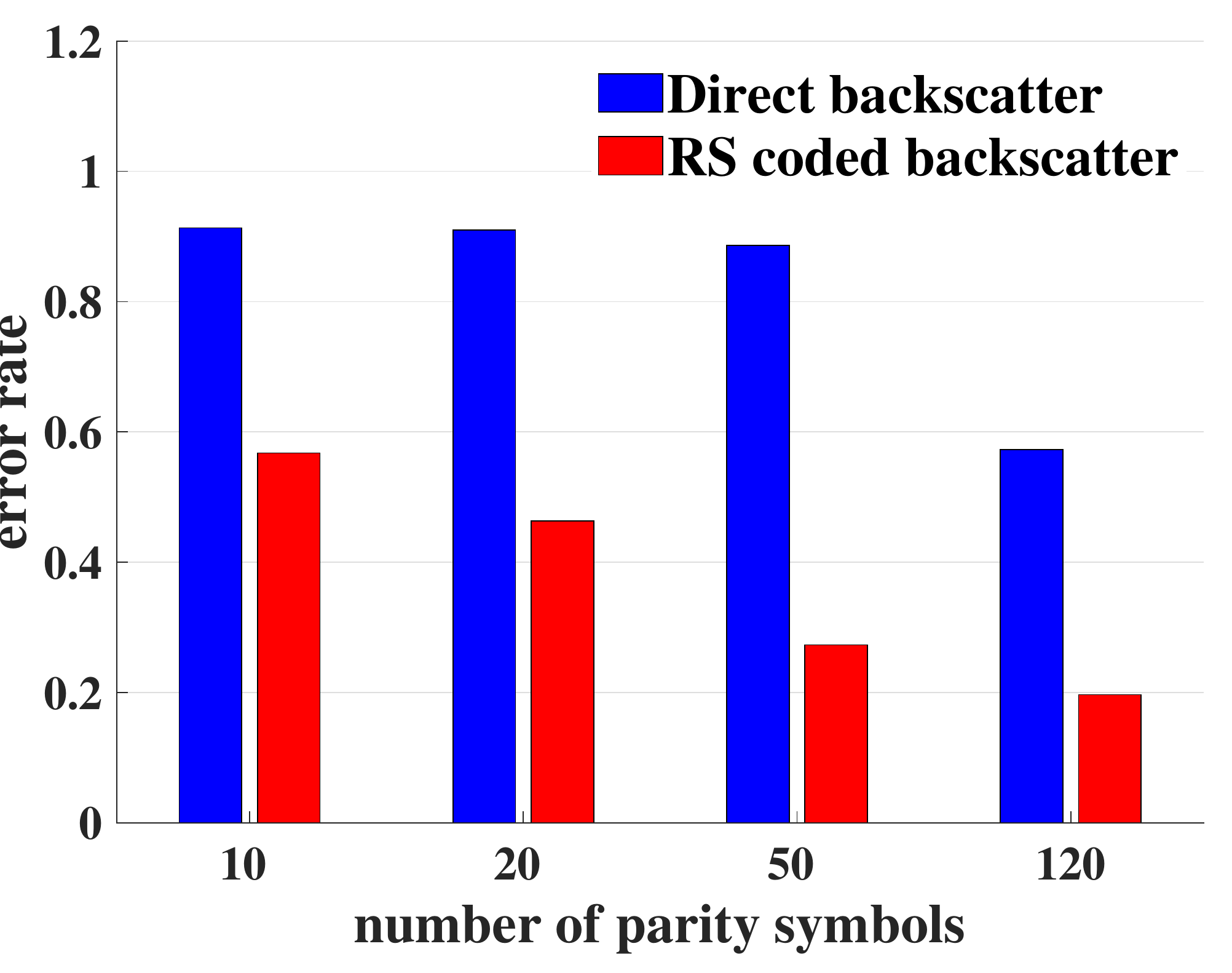}}
  ~
  \subfigure[probability distribution]{
        \includegraphics[width=0.22\textwidth]{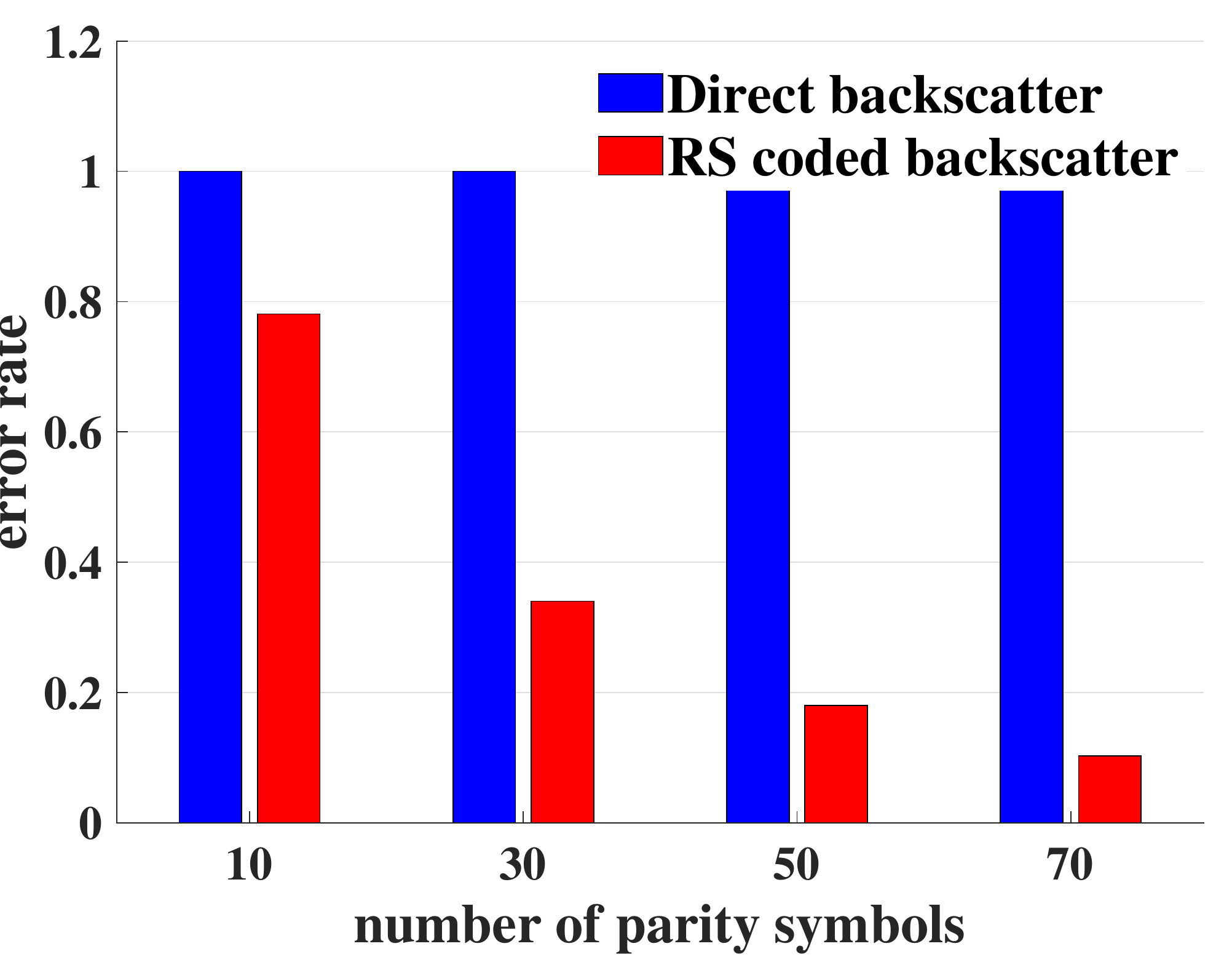}}
\caption{The effectiveness of adopting RS code for backscattering over intermittent WiFi traffics. WiFi signals in (a) and (b) are generated by a USRP device with different packet size, (c) is obtained by sniffing the channel 1 in our lab, (d) is generated from the probability distributions.}
\label{fig:RS_effect}
\end{figure*}

As with any error-correcting code, the principle of an RS code is to transform a given information sequence into a longer sequence called a codeword by adding some redundancy.  In this way the information can be recovered by performing a decoding process. A Reed-Solomon code is specified as RS($n, k$) with $m$-bit symbols. The RS encoder takes $k$ data symbols of $m$ bits each and adds parity symbols to construct an $n$ symbol codeword. There are $n-k$ parity symbols in a codeword for protecting the data symbols. RS codes are defined over symbols, i.e., over a Galois field (GF) of size $2^m$.  An RS code can correct $t = \lfloor \frac{n-k} 2 \rfloor$ symbol errors, and are optimal in this regard --- RS codes achieve the Singleton bound with equality. As shown in Fig.~\ref{fig:rs}, for GF(8), each symbol is represented by 3 bits. If one or more bits have an error, then the whole symbol is in error. Since the RS code corrects symbol errors, it does not matter if one or all 3 bits have an error $-$ it only counts as 1 out of $t$ symbols to be corrected. However, a 3-bit burst error could span up to two symbols, and in this case the RS code must correct 2 out of $t$ symbols. This is the reason RS codes work well for burst-error correction.

{\it RS encoding.} Let $\mathbf{\Phi}(n \times k)$ be a generator matrix of an RS($n,k)$ code.  If $\mathbf u$ is a $k \times 1$ vector of information symbols from $\textrm{GF}(2^m)$ (consisting of $k\cdot m$ bits), then $\mathbf c = \mathbf \Phi \cdot \mathbf u$ is an $n \times 1$ codeword of symbols from $\textrm{GF}(2^m)$ (consisting of $n \cdot m$ bits). A matrix $\mathbf \Phi$ such that any $k$ rows are linearly independent exist, and we are interested in the case that $\mathbf \Phi$ generates a systematic code.  That is, 
$k$ input data words are explicitly present among the $n$ generated codewords. That is, $\mathbf{\Phi}_{\text{sys}} = [\mathbf{I}_k | \mathbf{\Psi}_{k,n-k}]$ and the information sequence $\mathbf u$ appears in the first $k$ positions of the codeword $\mathbf c$.

\textit{RS encoding complexity.} We are primarily concerned with the encoding complexity, which takes place at the tag. While the Vandermonde form for the RS generator matrix is well known, recently it has been shown that the Hankel matrix has lower complexity when encoding systematic RS codes \cite{mattoussi2012complexity}. This approach may be suitable when the RS code parameters are dynamic.  Alternatively, the parity portion of the generator matrix $\mathbf{\Psi}$ may be generated once and stored. The storage requirement is $k \cdot (n-k)$, and the computational requirement is also proportional to $k \cdot (n-k)$, with mathematical operations over the GF($2^m$).

{\it RS decoding.} If $t$ or fewer symbol errors occur during transmission, then an RS decoder can recover the transmitted RS codeword and the transmitted data is in positions $1$ to $k$ of this codeword.  The design of efficient decoding algorithms has been the study of intense research for years, and now both hardware and software implementations are widely used in practice.  The best known of these is the Berlekamp-Massey decoding algorithm, which finds the roots of an error-locator polynomial.  While the complexity of the most efficient algorithms is proportional to $n \log n$, this operation is performed at the reciever, and does not contribute to the tag's computational burden.

\section{Reliable and Simple Backscatter Communication}
In this section, the proposed algorithms for achieving sustainable backscatter communications which ride over WiFi signals is presented.

\subsection{Why do we need to adopt RS codes in rider mode?}
In rider mode, it is hard to control the excitation signals for backscattering. In addition, the silent duration is existing and varying in the exictation traffics, the performance of backscatter communicaitons is significantly affected. As the backscattered information bits from the tag are consecutively lost within the off states, the backscatter link can be viewed as a burst error channel. Based on the preliminary knowledge of RS codes, we aim to examine whether adopting RS codes is an effective solution in backscatter communication to improve the reliability. To answer this question, some preliminary results are obtained by performing Monte Carlo simulations for different sets of time duration of on and off states. It is counted as an error in baseline systems when the transmitting time of a frame exceeds the duration of an on state, and in RS coded backscatter when the lost symbols exceeds the maximal error correcting capability of RS codes. We perform 5000 simulations to evaluate the average error rate for both systems.

The results in Fig.~\ref{fig:RS_effect} are obtained by fixing the length of a codeword $n=127$ and adjusting the number of parity/redundant symbols in the RS code. The excitation WiFi signals for obtaining the results of Fig.~\ref{fig:RS_effect} are generated by a USRP device in Figs.~\ref{fig:RS_effect} (a) and (b), real WiFi traffic from a TP-Link AP in Fig.~\ref{fig:RS_effect}(c) and the Pareto distributions. It is obvious that adopting RS code significantly reduce the error rate of the backscatter communications in different scenarios because of its strong ability in correcting burst errors. Furthermore, RS code is suitable for small size of data, which is typical for backscatter communications, and its encoding complexity is proportional to $k (n-k)$. Therefore, adopting RS code in rider mode is effective and feasible.

\subsection{Why do we need to optimize RS code for backscattering?}
As stated above, RS shows its effective improvements of the error rate in the context of riding over the intermittent signals. It is found from Fig.~\ref{fig:RS_effect}(a)-(c) that the error rate converges to a certain level when the number of redundant symbols is relatively enough, and no more gain will be obtained even when the number of redundant symbols is increasing. In other words, it wastes bandwidth of backscatter links if we add too many redundant symbols (less information is backscattered within a frame). In short, we {\it must} know how many parity symbols should be added to protect the backscatter communication in different scenarios to achieve a good balance between the efficiency and reliability. Thus, we propose an optimization algorithm to search an optimal RS code to maximize the efficiency while satisfying the reliability. The proposed optimization algorithm consists of three main steps: (1) estimating the probability distribution for on and off duration of WiFi traffics, (2) constructing a Markov model to describe the transition between the states, (4) optimizing the RS code to meet the constraint on reliability.

\subsection{Flow-oriented estimation.}
The RS code design highly depends on the WiFi traffic flow from the excitation source. Intuitively, if the WiFi packets are highly occupied in the channel, the redundancy of the RS code can be short. Conversely, more redundant symbols have to be added in order to backscatter information excited by less packets. Hence, in order to design an adaptive RS code, the tag have to know the states of excitation traffics, at least, the statistical knowledge. However, due to the limited power of the tag, it is infeasible to optimize the code at the tag. We thereby shift the optimization process from the tag to the receiver in our system and feedback an index indicating the optimal RS code parameters to the tag. Specifically, the receiver first listen to the desired channel and measure the duration of the on and off states based on the received signals, respectively. The measurements are then input into an MLE-based estimator to generate the probability distribution of each state.

{\it Measurement.} The receiver sniffs the channel and evaluates the duration of on and off states of WiFi traffics on a desired channel, e.g, channel 1. The observed data sequences are stored as $\mathbf{x}^N=[x_1, x_2, \cdots, x_N]$ and $\mathbf{y}^M=[y_1, y_2, \cdots, y_M]$ representing the duration of the on and off states in microsecond, respectively. The lengths $N$ and $M$ does not required to be identical.

{\it Estimation of probability distribution.}
The measurement data is then adopted to obtain the parameters of probability distribution for WiFi traffic. For simplicity, we use the Pareto distribution \cite{Pareto} to evaluate the statistical property of the WiFi traffics, both for the states of WiFi signal being on and off.
Based on the Pareto distribution, the probability density function ({\it pdf}) representing the off state is
\begin{equation}\label{equ:pareto}
  f_{\tau}(\tau; \lambda) = \begin{cases}
  \lambda \frac{\tau_m^\lambda}{\tau^{\lambda+1}}, \quad  &\tau\ge\tau_m \\
  0, & \tau<\tau_m
  \end{cases}
\end{equation}
where $\tau$ represents the silent duration, of which the possible minimum value is denoted by $\tau_m$, and $\lambda$ is a positive shape parameter of the Pareto distribution. Similarly, the {\it pdf} of on states can be expressed with parameters $\mu, \delta$ and $\delta_m$, which represent the shape parameter of the Pareto distribution, durations of on states and its minimum value, respectively.

The MLE method is utilized to estimate the unknown parameters of the Pareto distributions from $\mathbf{x}^N$ and $\mathbf{y}^M$, respectively. Let $\mathbf{\theta}$ denote the parameters ($\tau_m, \lambda$) to be determined. The likelihood function of the Pareto distribution given samples $\mathbf{x}^N$ is
\begin{equation}\label{equ:likelihood}
  \mathcal{L}(\theta; x) = \prod_{i=1}^{N} \lambda\frac{\tau_m^\lambda}{x_i^{\lambda+1}},
\end{equation}
of which the log-likelihood function is
\begin{equation}\label{equ:llr}
  \log[\mathcal{L}(\theta; x)] = N\log(\lambda) + N\lambda\log(\tau_m)-(\lambda+1)\sum_{i=1}^{N}\log(x_i),
\end{equation}
after several elementary steps of calculations.

The estimation of $\hat{\theta}$ is obtained by maximizing $\log[\mathcal{L}(\theta; x)]$ in \eqref{equ:llr}. By performing the partial derivative of $\log[\mathcal{L}(\theta; x)]$ with respect to $\lambda$ and set it to zero, we obtain the optimal estimation $\hat{\lambda}$ for the shape parameter $\lambda$ as
\begin{equation}
\label{equ:lambda}
\hat{\lambda}=\frac{N}{\sum_{i=1}^{N}\log(x_i)-N\log(\hat{\tau}_m)}.
\end{equation}
However, it is an unbounded problem for estimating $\tau_m$ since the log-likelihood function is monotonically increasing with respect to $\tau_m$. Taking the physical meaning of $\tau_m$ which represents the possible minimum value of $\tau$ into account, the estimation $\hat{\tau}_m$ is simply obtained by choosing the minimum from $\mathbf{x}^N$ as $\hat{\tau}_m = \min\{x_1, x_2, \cdots, x_N\}$.

Similarly, we obtain the estimated parameters for on state, as
\begin{equation}\label{equ:mu}
  \hat{\mu}=\frac{M}{\sum_{i=1}^{M}\log(y_i)-M\log(\hat{\delta}_m)}.
\end{equation}
with $\hat{\delta}_m$ being the minimum of $\mathbf{y}^M$.

\subsection{Symbol-oriented estimation.}
\begin{figure}
  \centering
  \includegraphics[width=2in]{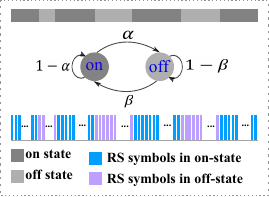}
  \caption{The two-state Markov model of burst error channel.}\label{fig:GEmodel}
\end{figure}
In the previous steps, the statistical information for on-and-off states of WiFi signals are obtained. We now turn to analyze the error rate of backscatter communication which rides on WiFi signals using RS code.

Inspired by the method presented in \cite{Yee95, Wolf98}, a two-state Markov model which is illustrated in Fig.~\ref{fig:GEmodel} is adopted here to represent the transition among the states. It should be emphasized here that the Markov model is the key for designing RS codes over the statistical intermittent channel. Let $\alpha$ and $\beta$ denote the transition probabilities from on state to off state, and off state to on state, respectively. The transition probability matrix $\mathbf{Q}$ of Markov model is thus given by
\begin{equation}\label{equ:transmatrix}
  \mathbf{Q} = \begin{bmatrix}
 1-\alpha& \alpha \\
 \beta & 1 - \beta
\end{bmatrix},
\end{equation}
then the average symbol error rate is obtained as \cite{Yee95}
\begin{equation}
p_s = \frac{\alpha}{\alpha+\beta}
\label{equ:ps}
\end{equation}
by assuming the symbol error only occurs when it experiences the off state. Hence, the symbol error rate $p_e$ of adopting an $(n, k)$ RS code can be calculated based on $p_s$ and the error-correcting capability, as \cite{Wolf98}
\begin{equation}\label{equ:ser}
  p_e = \sum_{i=t+1}^{n} \binom{n}{i}p_s^ip_s^{n-i}.
\end{equation}

The error rate $p_e$ can be obtained if parameters $\alpha$ and $\beta$ are given, hence we need to estimate them. By using the statistical knowledge of the Pareto distributions for on-and-off states, the estimation of $\alpha$ and $\beta$
\begin{equation} \label{equ:markov}
\begin{cases}
\hat{\alpha} =  R / \bar{\delta} \\
\hat{\beta}  = R / \bar{\tau}
\end{cases},
\end{equation}
where $\bar{\tau}$ and $\bar{\delta}$ are the mean durations of on and off states of WiFi signals, and $R$ is the transmission rate of backscatter link. By substituting \eqref{equ:markov} into Equations \eqref{equ:ps} and \eqref{equ:ser}, we can obtain the estimation on the error rate $p_e$, which is related to the reliability of the backscatter system.

\subsection{RS code optimization.}
Our aim is to construct an RS code such that the transmission efficiency is maximized and the backscatter data can be recovered over the intermittent WiFi signals. The efficiency is represented by the coding rate $\frac{k}{n}$ of RS codes, i.e., higher the coding rate, more delivered information within a frame, and vice versa. Hence, the optimization problem for constructing RS code can be represented as $\max k/n$ such that $p_e\rightarrow0$. However, it is difficult to handle it in practice because it is very time-consuming to search the entire set of RS codes. Given the fact that the tag has limited information to be transmitted, and the encoding complexity of RS codes increases as $n$ increases, typically the encoding complexity is $O(n\log n)$. Hence, we transfer the optimization problem of RS codes by searching a given subset of RS codes, as
\begin{figure*}
  \centering
  \includegraphics[width=6.5in]{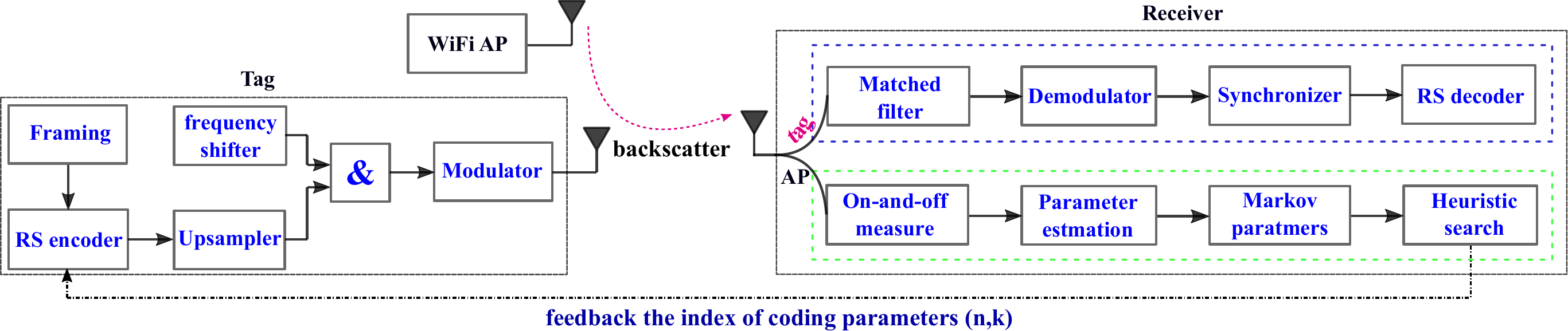}
  \caption{The main processes of the tag and the receiver of the proposed RS-coded backscatter system GuardRider.}
  \label{fig:sysmodel}
\end{figure*}
\begin{equation}\label{equ:max}
  \max_{(n, k)} \quad \frac{k}{n}
\end{equation}
such that
\begin{align}
   p_e &\le p_e^{th}, \label{equ:con1}\\
   \log_2(n+1) &\in \mathbb{Z} \label{equ:con3} \\
   k / 2 &\neq 0, \label{equ:con4} \\
   n &\in \{7, 15, 31, \cdots, 127\}, \label{equ:con2}
\end{align}
where $p_e^{th}$ is a predefined threshold for the symbol error rate\footnote{The symbol error rate is not equivalent to the BER since only 1 erroneous bit within a symbol leads to an error symbol.} and $\mathbb{Z}$ denotes the set of all integers.Please be noted that using threshold $p_e^{th}$ is to accelerate the speed of the convergency of the algorithm. Otherwise, if let $p_e^{th}$ be $0$, a large $p_s$ may cause bad convergency of the algorithm. The constraint \eqref{equ:con1} is deriving from the symbol error rate of adopting an $(n,k)$ code, while the constraints \eqref{equ:con3}-\eqref{equ:con4} are from the requirement of the RS code. The constraint \eqref{equ:con2} limits the searching domain. It is obvious that the optimization problem of \eqref{equ:max} is not convex since the objective function is an affine divided by an affine. In order to find the optimal code parameters $(n,k)$ within the subset, we propose a heuristic searching algorithm to select the RS code parameters such that the error rate is below $p_e^{th}$ and the efficiency is maximized. The heuristic searching algorithm is summarized in {\bf Algorithm} 1.

\begin{algorithm}[t!]
	\small{
		\caption{Heuristic search of the RS code}
		\label{alg:LPF_filter}
		\begin{algorithmic}[1]
			\Require $\mathbf{x}^N$, $\mathbf{y}^M$, $R$, $P_e^{th}$
			\Ensure  $n$, $k$
             \State $\tau_m \gets \min\{\mathbf{x}^N\}$;
             \State $\lambda \gets $ the estimation result using Equation (\ref{equ:lambda});
             \State $\delta_m \gets \min\{\mathbf{y}^M\}$;
             \State $\mu \gets$ the estimation result using Equation (\ref{equ:mu});
             \State $\bar{\tau} \gets \lambda\tau_m/(\lambda-1)$; $\bar{\delta} \gets \mu\delta_m/(\mu-1)$;
             \State $\alpha \gets R/\bar{\delta}$; $\beta\gets R/\bar{\alpha}$; $E\gets0$;
             \For {$m \gets 3; m \le 7; m++$}
                 \State $\tilde{n} \gets 2^m - 1$;
                 \For {$j\gets\tilde{n}-2; j \ge 1; j=j-2$}
                 \State $t \gets (\tilde{n}-j)/2$;
                 \State $p_e \gets$ the obtained result using Equation (\ref{equ:ser});
                 \If {$p_e \le P_e^{th}$}
                 \State $\tilde{k} \gets j$; break;
                 \EndIf
                 \EndFor
                 \If {$\tilde{k}/\tilde{n} > E$}
                 \State $n\gets\tilde{n}$; $k\gets\tilde{k}$; $E\gets\tilde{k}/\tilde{n}$;
                 \EndIf
            \EndFor
		\end{algorithmic}
	}
\end{algorithm}

\section{System Design and Implementation}
Figure~\ref{fig:sysmodel} demonstrates an overview of our system GuardRider. A commercial WiFi router transmits regular WiFi signals packet-by-packet as the excitation, the tag reflects the excitation signals to a receiver while conveying its information. The tag shifts the frequency of the reflected signal to another channel by the frequency shifter, to avoid the interference. The receiver listens to the shifted frequency band and decodes the information based on the received power. The fundamental components of our system which enables backscatter information is detailed below.
\subsection{Framing and RS encoding}
As shown in Fig.~\ref{fig:sysmodel}, the tag in the backscatter system performs framing, RS encoding, baseband conversion, upsampling and frequency shifting to backscatter its information.

The information sequence represented by bits are first processed by the Framing block to form a data frame, where the frame structure consists of the following fields: 
\begin{itemize}
\item $1$~byte length to indicate the length of the frame.
\item $3\sim108$~bytes payload data.
\item $2$~bytes cyclic redundancy check used to detect any intransit corruption of data.
\end{itemize}

Every $m=\log_2(n+1)$ bits in the data frame are grouped together and converted to one symbol in GF($2^m$). After that, every $k$ symbols are encoded to an $n$-symbol codeword based on the optimal parameters $(n, k)$ obtained by {\bf Algorithm} 1. The encoded symbols are finally converted back to a bit sequence is then converted using non-return-to-zero(NRZ) line code.

For frame synchronization, we insert a preamble sequence of $\{101010101010101010101010110100100011\}$ before the baseband signal, similar to the FM0 coding \cite{EPC}.
\subsection{Frequency shift}
For simplicity, the excitation signal can be represented as a sinusoidal signal, denoted as $\sin(2\pi f_c t)$. The micro-controller of the backscatter tag generates a square wave with frequency $\Delta_f$ to switch the states between absorbing and reflecting. Since the first harmonic of a square wave is also a sinusoid signal based on the well-known Fourier analysis, when applying to incoming sinusoidal signal, a frequency shift $\Delta_f$ is achieved in the backscatter link. Thus, we can simplify the process of square wave to a sinusoid signal using the first harmonic, as $\sin(2\pi \Delta_f t)$. Consequently, the process of backscattering can be represented by the product of the aforementioned two sinusoidal signals, which is given by $ 2\sin(2\pi f_c t)\sin(2\pi \Delta_f t) = \cos(2\pi(f_c-\Delta_f)t)-\cos(2\pi(f_c+\Delta_f)t) $. As a process for the received backscattered signals, the receiver turns its center frequency to one of the shifted signal, which is $ f_c-\Delta_f $ or $ f_c+\Delta_f $, to receive the signals in the backscatter link. The purpose of the frequency shift is to eliminate the interference from the main link to the backscatter link, as the signal strength of the main link is significantly stronger than that of the backscatter link.

An upsampling process is then performed to increase the frequency of the baseband signal in order to match the shifting frequency, followed by a logic AND operation between the upsampled signal and the square wave with frequency $\Delta_f$. The output signal from the AND operator controls the antenna to backscatter information using OOK modulation. 

\subsection{Receiver}
After detected a frame by examining the autocorrelation value of the received samples, the receiver starts to taking the in-phase (I) and quadrature (Q) samples for each transmitted symbols with a sampling frequency $f_s$~MHz. Let $r_I$ and $r_Q$ represent the obtained I/Q samples, and the power of the samples is calculated by $P=\sqrt{r_I^2+r_Q^2}$. The power sequence of the samples is then pass through a matched filter. The demodulator performs in the following manner to obtain the decisions, as
\begin{equation}\label{eq:detection}
  \hat{c} = \begin{cases}
  1 \quad \text{if} \quad P \ge P^{th}, \\
  0 \quad \text{otherwise},
  \end{cases}
\end{equation}
with $P^{th}$ being a detection threshold, which is adaptively adjusted based on the received power strength. We performs cross-correlation process between the known preamble pattern and the received samples. Once the peak is detected among the cross correlation values, of which the position is considered as the end point of the preamble. The threshold $P^{th}$ is then set at the average of the minimum representing ``1'' and the maximum representing ``0'' in the samples. It should be mentioned here that the reason of performing adaptive adjust is that the received power strength varies considerably in the context of backscatter communications.

After that, a downsampling process is executed to filter the bit samples. However, the timing synchronization between the tag and the receiver should be properly established. Otherwise, the error probability is significantly increased with inferior timing recovery. At the receiver, the symbol synchronizer based on interpolation filter, zero-crossing timing error detector and modular-1 interpolation control is implemented before the decoding process. By adopting the synchronizer, we properly take the RS coded bit sequence with burst errors, finally the RS decoding process is carried out to obtain the data frame.  

\subsection{Implementation}
We build a prototype of GuardRider using a commodity WiFi transmitter, a customized FPGA-based tag and a USRP receiver.

{\bf WiFi transmitter:} The WiFi transmitter is a TP-Link router, which supports 802.11a/b/g/n/ac protocols.
For evaluating the performance of the RS code against the intermittent WiFi siganl, we use a laptop to run {\it ping} command to generate request to the WiFi router with different packet size and interval.

{\bf Backscatter Tag:} The backscatter tag is connected to an NI USRP 2953R with FPGA through digital input/output (DIO) cable. The transmitted process of the tag node is implemented at the FPGA side, and then the square wave which carries the information is transmitted to the tag. 

The prototype of the tag consists of an antenna and a micro-controller, which controls the single pole double throw (SPDT) RF switch network to generate backscatter signals.

{\bf Receiver:} The receiver performs two tasks in GuardRider, which are optimization process and receiving backscatter information. The function of each block is implemented based on Labview platform. 

{\it Optimizaiton.} A USRP RIO device first listens the channel of the legacy link and takes the in-phase (I) and quadrature (Q) samples of the channel for the legacy link with sampling frequency 5MHz. The signal power is then calculated using I/Q samples and the durations of on and off states are calculated. The durations are finally fed into the optimization algorithm containing three blocks to obtain the optimal RS code parameters.
\begin{figure}
  \centering
  \subfigure[Schematic diagram of FPGA implementation of the tag]{
        \includegraphics[width=0.25\textwidth]{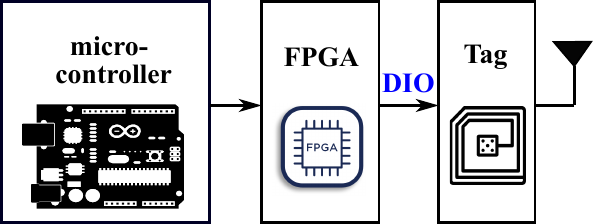}
  }
  ~
  \subfigure[Hardware prototype]{
        \includegraphics[width=0.15\textwidth]{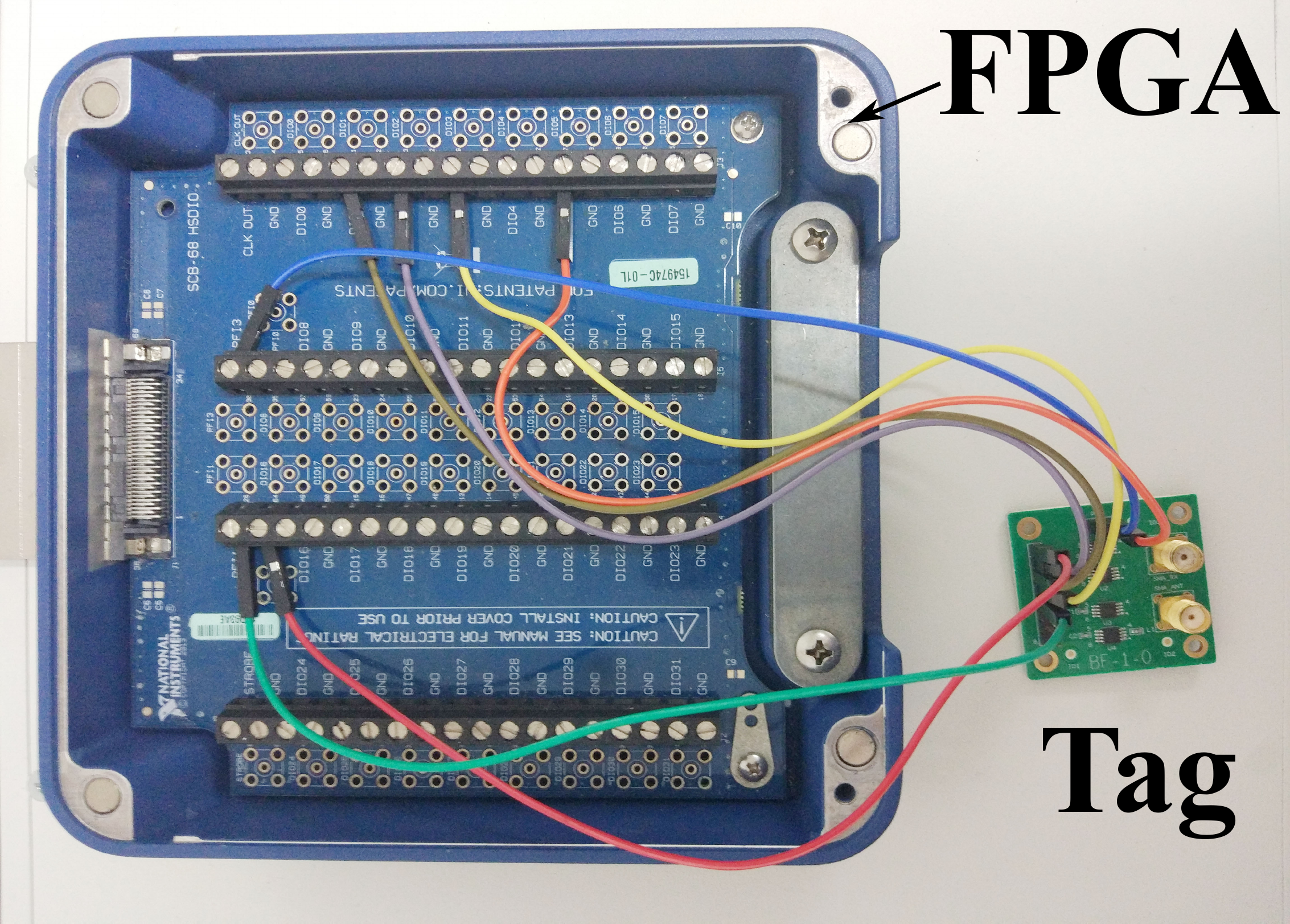}
  }
  \caption{The FPGA implementation of our customized tag.}
  \label{fig:tag}
\end{figure}

{\it Receive backscattered information.} The USRP RIO device is turned on to listen the channel at the shifted frequency. Once a frame is detected by comparing the auto-correlation of the received signal to a predefined threshold, the device starts to taking I/Q samples, followed by the matched filter process, demodulation, synchronization and RS decoding. 

\section{Performance Evaluation}

We evaluate the performance of GuardRider over WiFi signals with various silent periods, in terms of three metrics, including BER, FER and throughput. The results obtained from the experiments and simulations show that
\begin{itemize}
\item Both the BER and FER performances are reduced notably by GuardRider in different scenarios based on the conducted simulations. Particularly, the BER and FER could be reduced to $10^{-6}$ when the silent period is short enough.
\item The obtained results through the built hardware prototype further verify that GuardRider can reduce the FER by average $99\%$ and $60\%$ in different scenarios.
\item The throughput performance of the GuardRider is improved notably since the baseline system cannot work in typical scenarios.
\end{itemize}
\begin{figure}
  \centering
  \includegraphics[width=1.45in, angle=90]{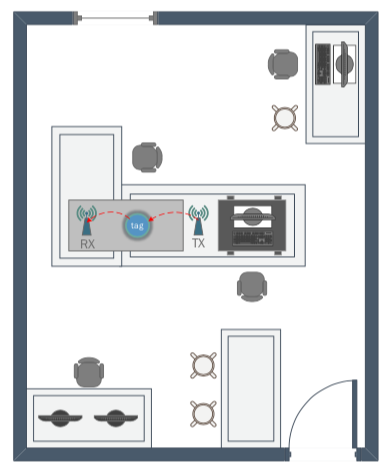}
  \caption{The floor plan of the office where the experiments are conducting.}\label{fig:floor}
\end{figure} 

\begin{figure*}
  \centering
  \subfigure[average silent duration $20\mu s$]{
        \includegraphics[width=0.3\textwidth]{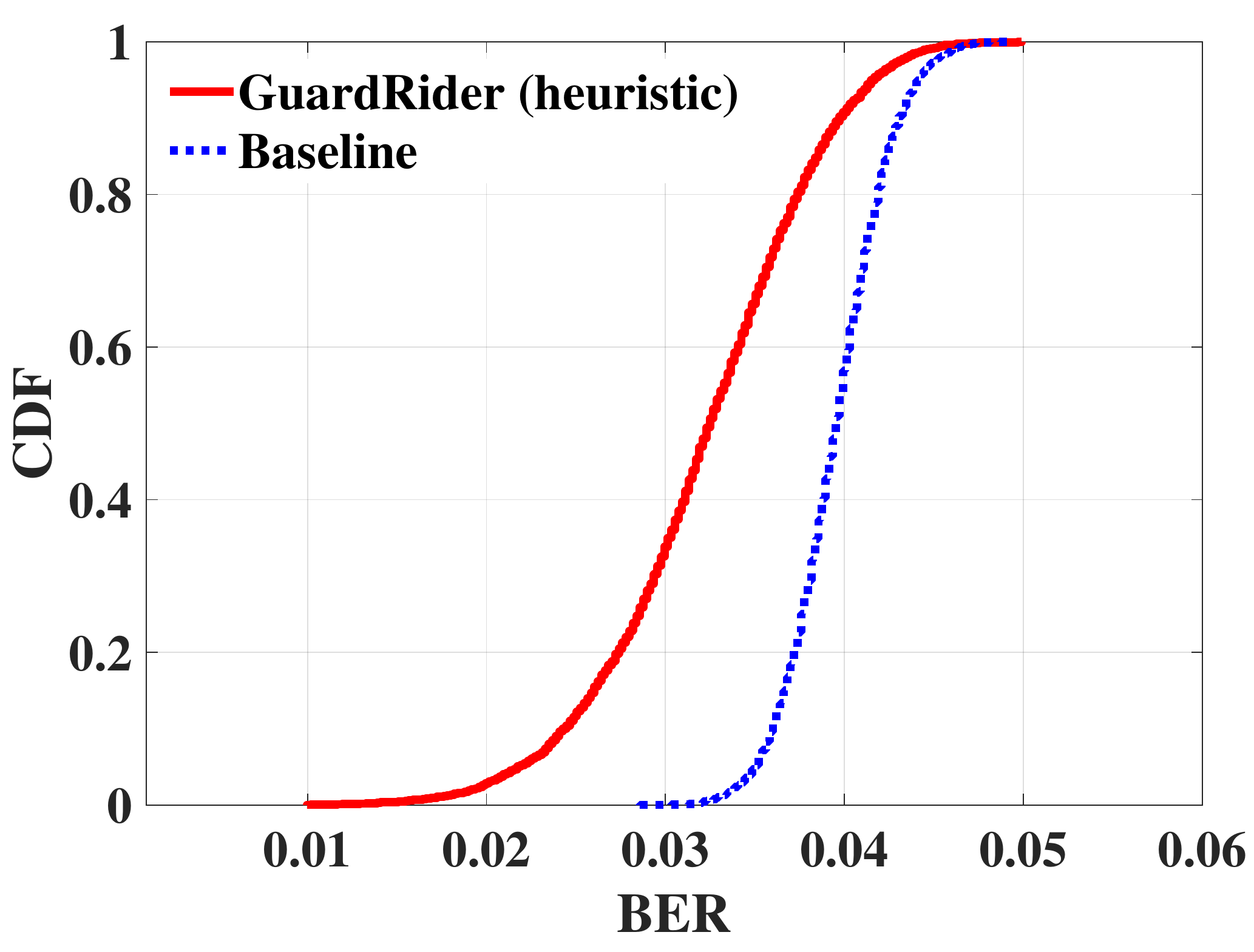}
  }
  ~
  \subfigure[average silent duration $40\mu s$]{
        \includegraphics[width=0.3\textwidth]{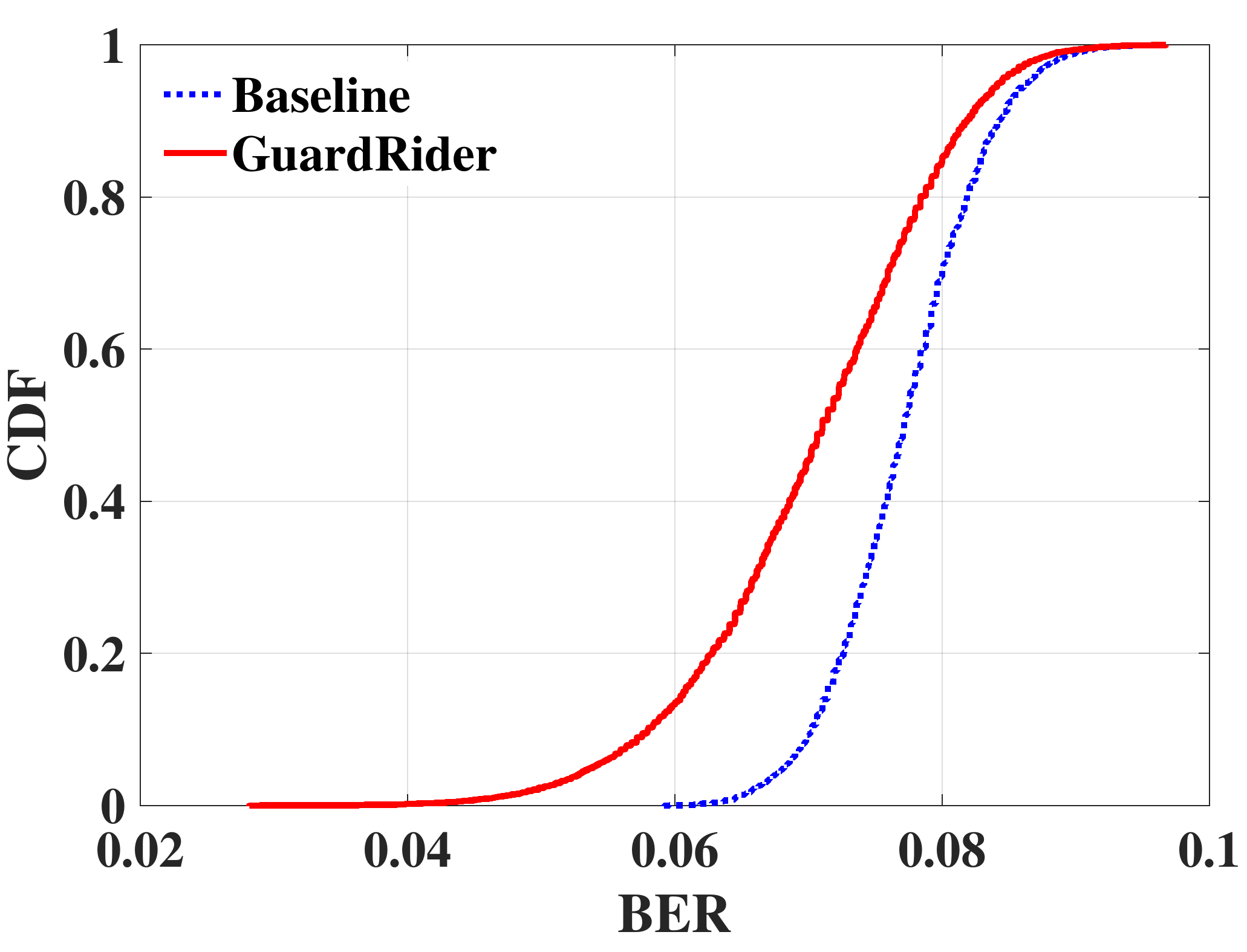}
  }
  ~
  \subfigure[average silent duration $60\mu s$]{
        \includegraphics[width=0.3\textwidth]{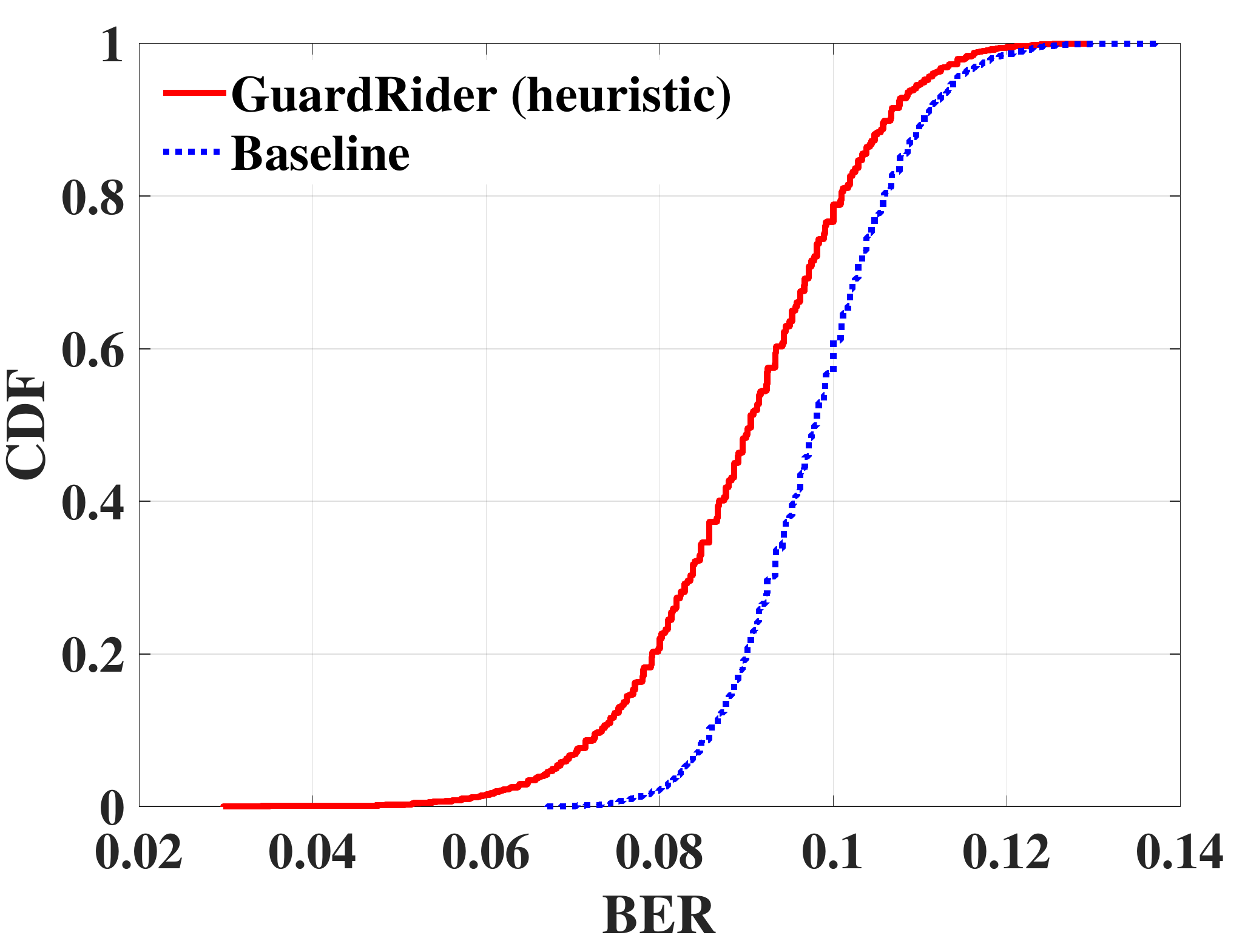}
  }
  \\
  \subfigure[average silent duration $20\mu s$]{
        \includegraphics[width=0.3\textwidth]{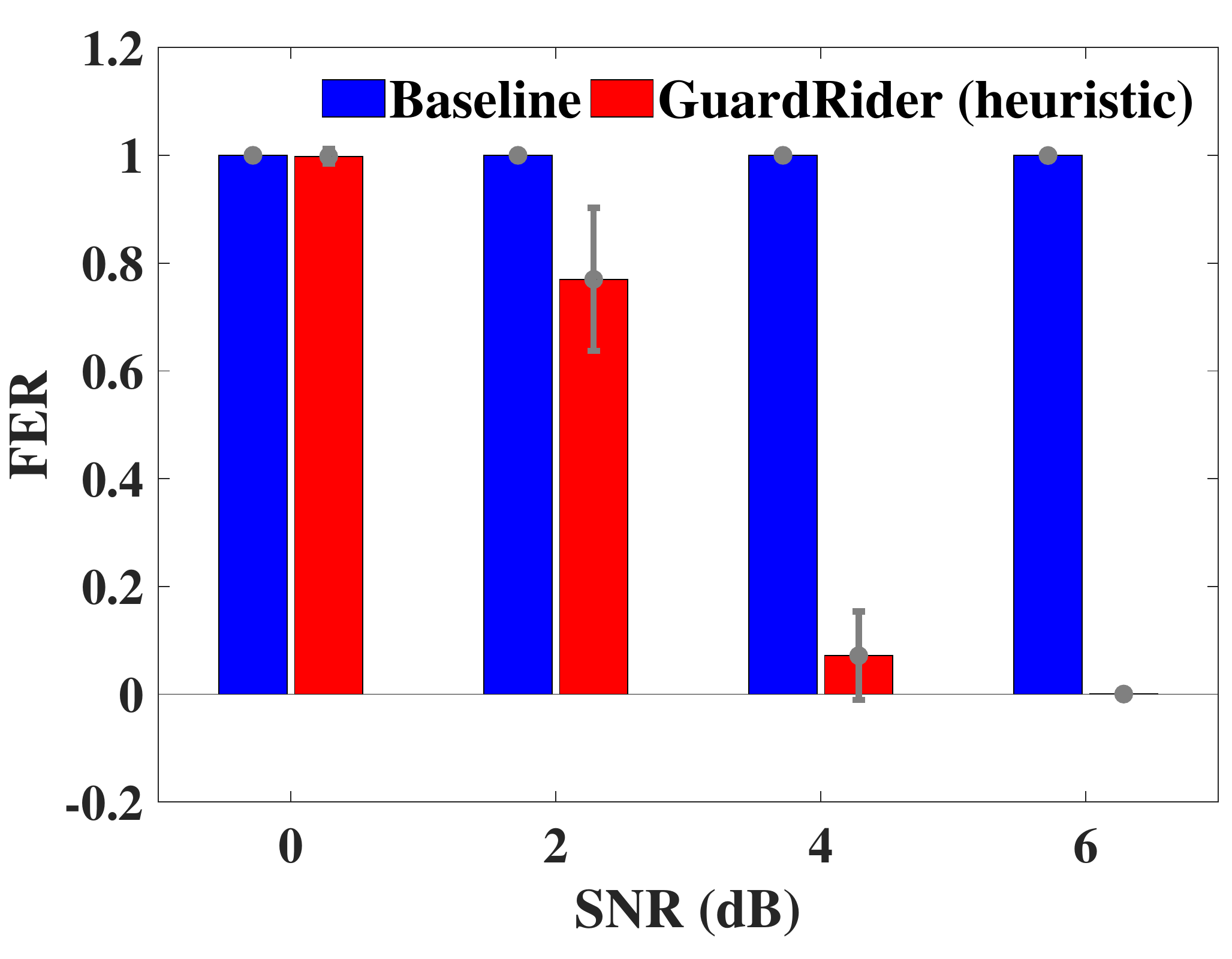}
  }~
  \subfigure[average silent duration $40\mu s$]{
        \includegraphics[width=0.3\textwidth]{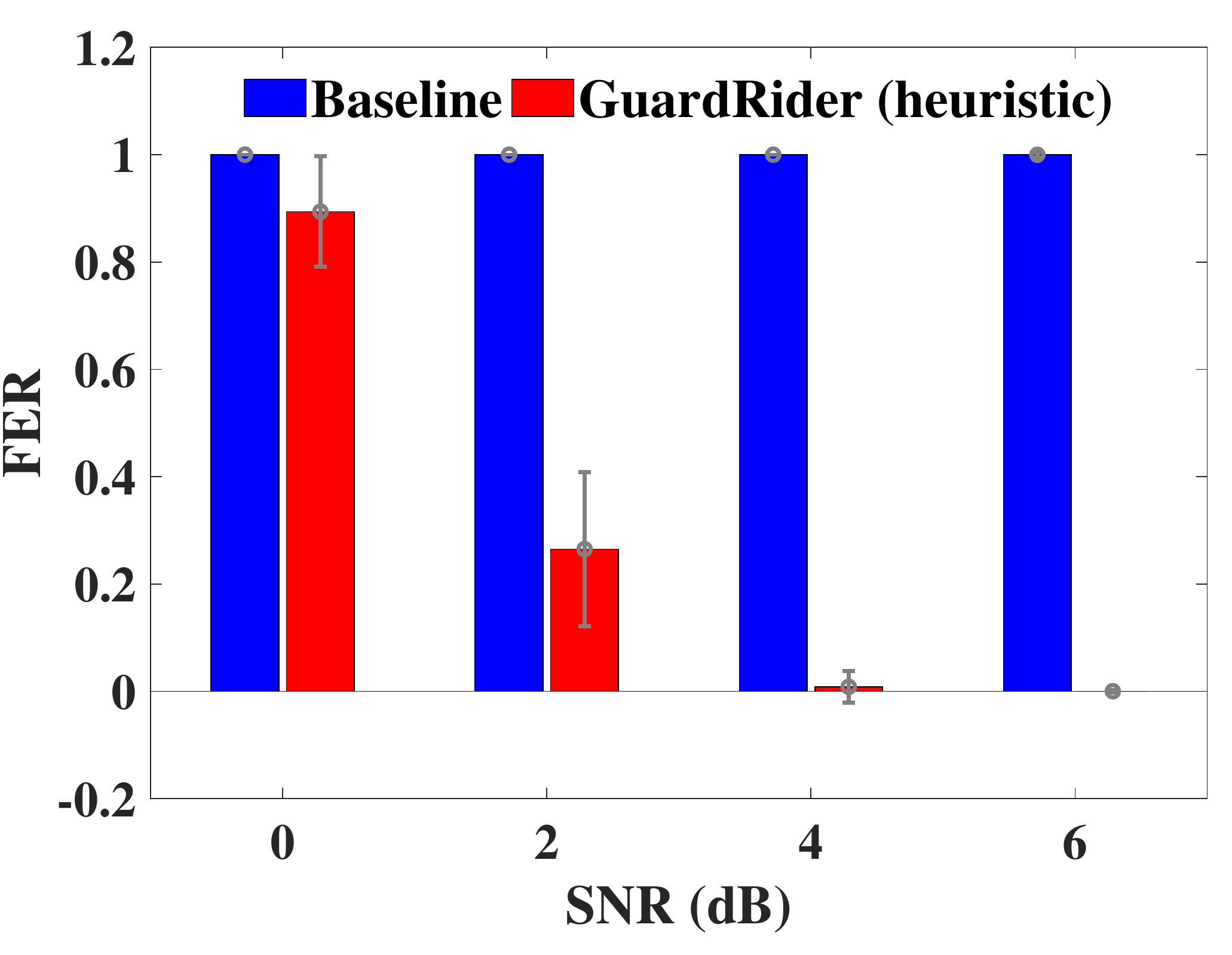}
  }
  ~
  \subfigure[average silent duration $60\mu s$]{
        \includegraphics[width=0.3\textwidth]{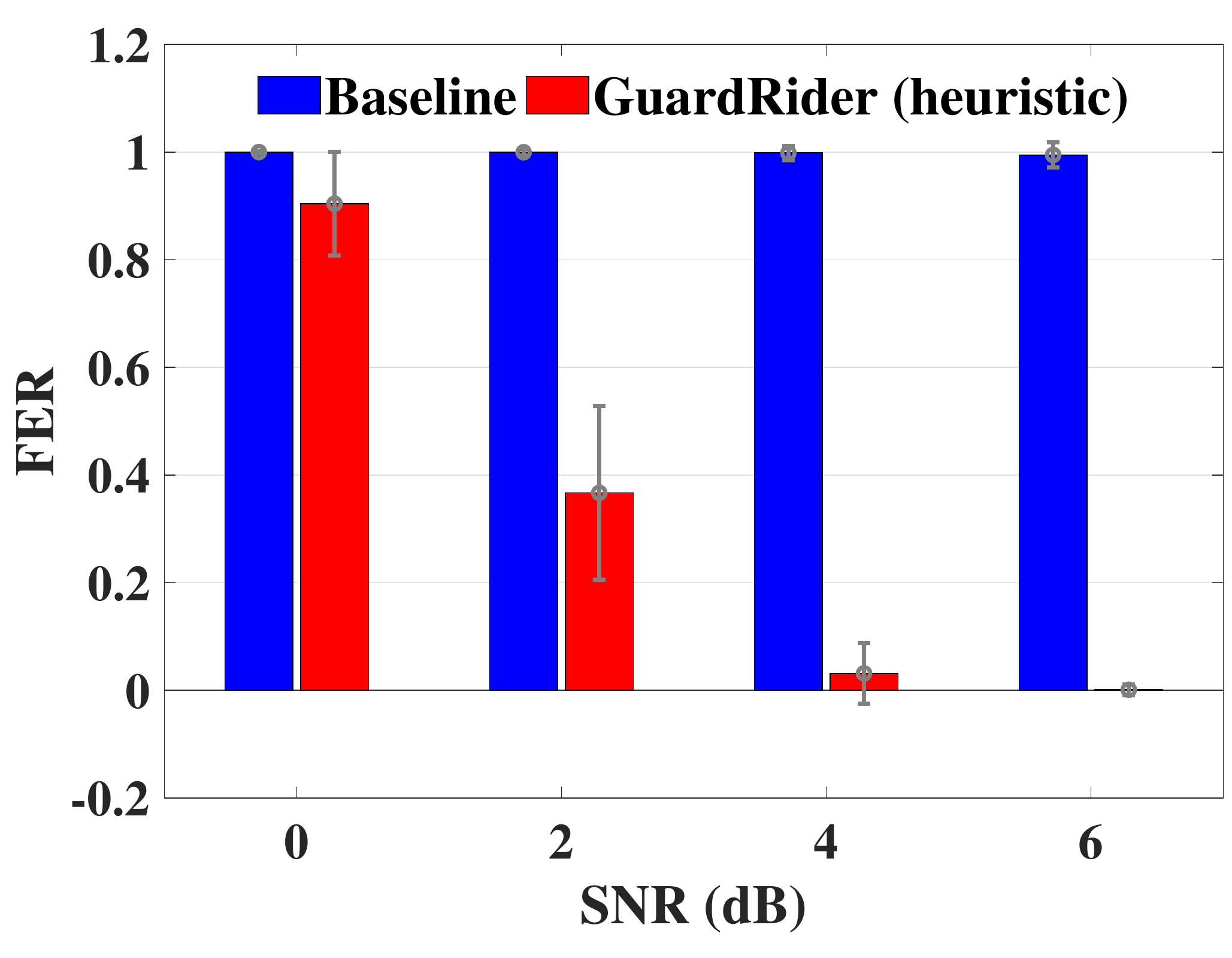}
  }
  \caption{The BER and FER comparison between the proposed GuardRider and baseline backscatter system based on simulations.}\label{fig:RS_ber}
  \label{fig:result3}
\end{figure*}
\subsection{Experiment setup}
We conduct our experiments in a $4 \times 6$m$^2$ office, of which the floor plan is shown in Fig.~\ref{fig:floor}. We use a Lenovo laptop which is 1m away from the WiFi AP to generate {\it ping} request with 2048 bytes and an interval $100\mu s$. The backscatter tag is located 0.3m and 0.6m away from the WiFi AP and the received antenna, respectively. We set the transmission speed at 6~Mbps and use channel 3 with a central frequency 2.422GHz and 20MHz bandwidth for the communication of the legacy link. Tag shifted the frequency by $50$MHz of the incoming excitation signal, i.e., the signal is shifted to channel 13 with a central frequency 2.472GHz, because of channel 13 is relatively clean compared to other channels and the interference from other WiFi networks can be ignored. Besides, the receiver set the $p_e^{th}$ at $10^{-3}$ for optimizaiton and the sampling frequency at 5MHz for taking the I/Q samples. 
\subsection{Impact of different silent period}
{\it Simulations.} We first perform a series of simulations to verify the impact of the traffic load to the BER and FER performance. We consider the following three cases:
\begin{itemize}
\item {\it Short silent duration.} The packet with average length 256 bytes are transmitted with an average silent duration $20$~$\mu s$.
\item {\it Medium silent duration.} The packet with average length 256 bytes are transmitted with an average silent duration $40$~$\mu s$.
\item {\it Long silent duration.} The packet with average length 256 bytes are transmitted with an average silent duration $60$~$\mu s$.
\end{itemize}
A total number of frames was $2000$ to evaluate the average BER and FER, where the results are depicted in Fig.~\ref{fig:RS_ber}. The threshold of $P_e^{th}$ were set at $10^{-3}$ in each case. It is found from Fig.~\ref{fig:RS_ber}, as the average duration of silent state decreases, GuardRider achieves much more gains compared to the baseline systems. In the scenario with short silent duration, GuardRider has hundredfold reduction both in BER and FER performance, while the improvements are relatively high in the case where the silent duration is quite long. 

\begin{figure*}
\centering
\subfigure[OFDM signals from USRP]{
        \includegraphics[width=0.275\textwidth]{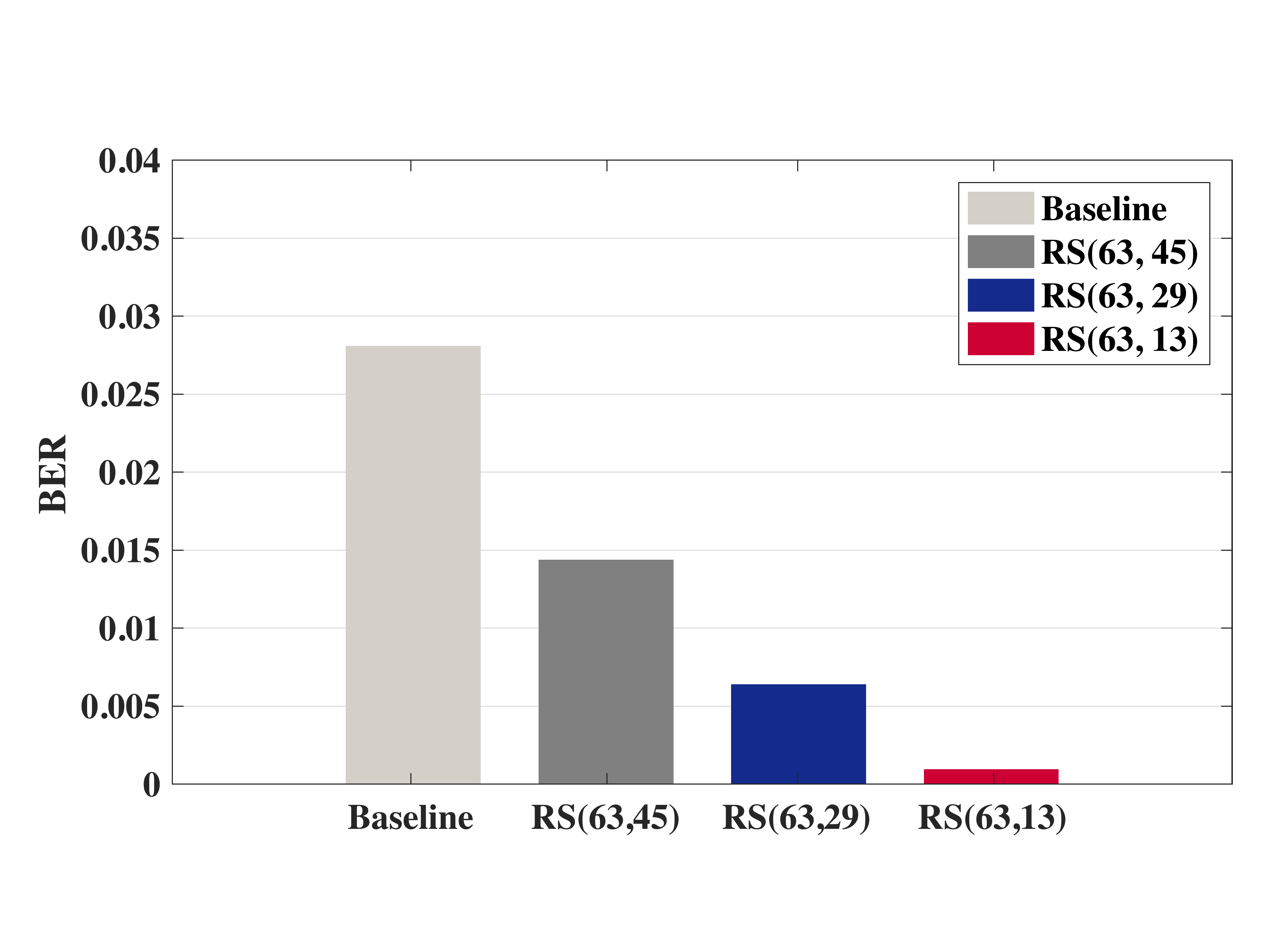}
  }
  ~
\subfigure[OFDM signals from USRP]{
        \includegraphics[width=0.275\textwidth]{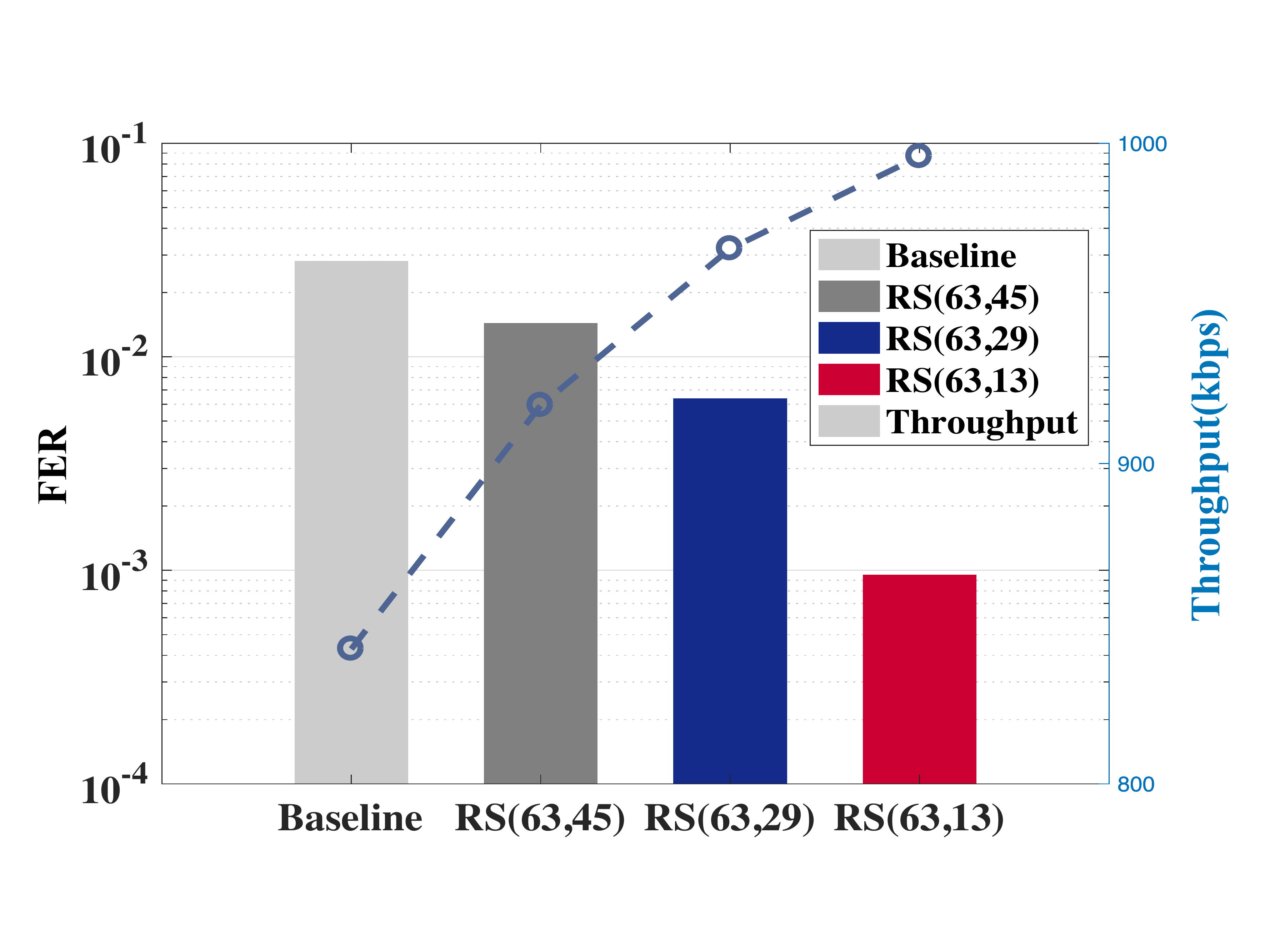}
  }
  \subfigure[WiFi traffic by ping]{
     \includegraphics[width=0.275\textwidth]{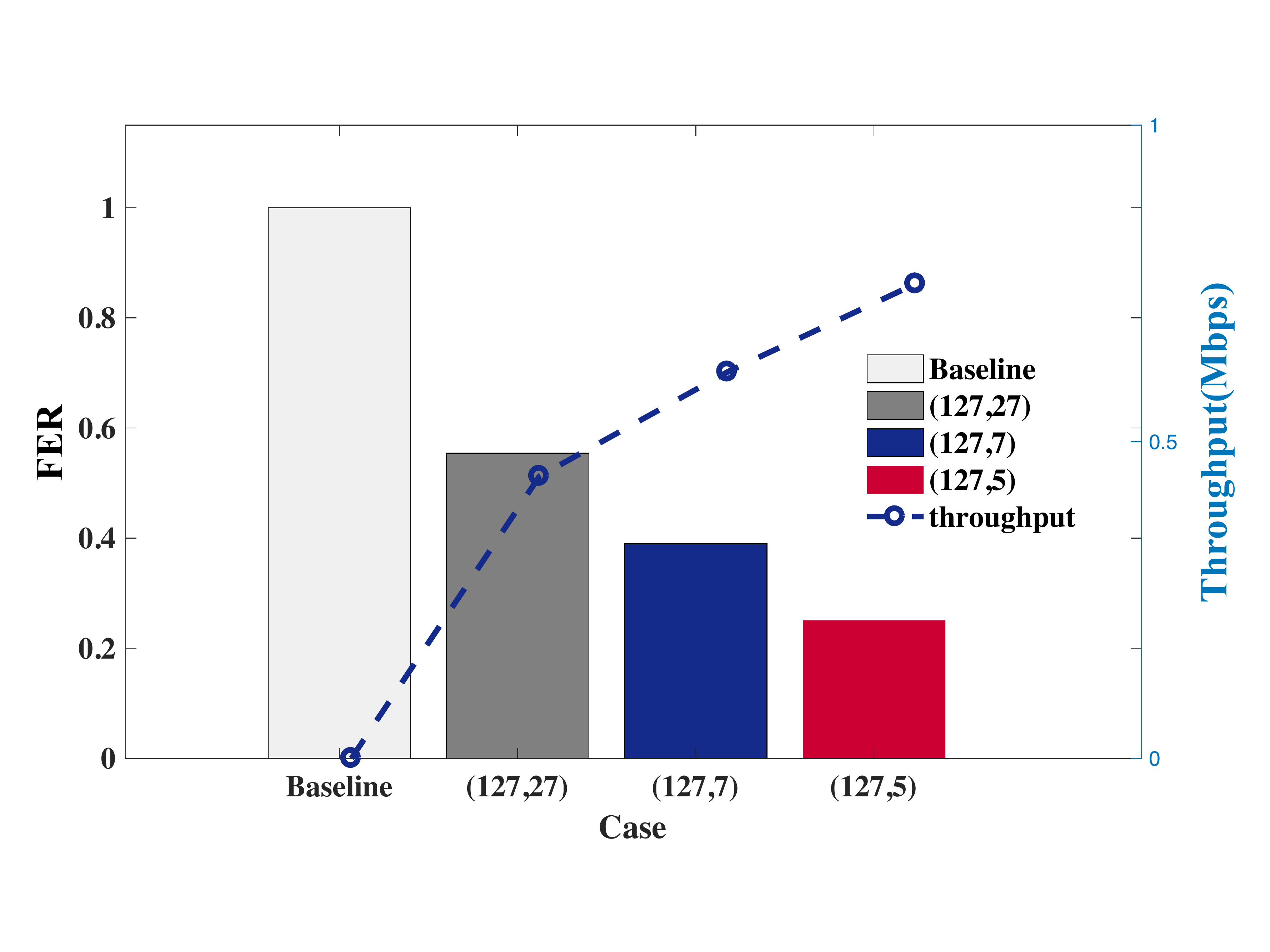}
     }
\caption{The performance of GuardRider excited by WiFi signals with frequency $2$GHz from USRP and $2.422$GHz from the WiFi AP.}
\label{fig:fer}
\end{figure*}

{\it Experiment.} Figure~\ref{fig:fer} illustrates the system performance in terms of BER and FER by using the built hardware prototype, where GuardRider adopts RS$(63, 45)$, RS$(63, 29)$, and RS$(63, 13)$ codes to protect the backscatter transmission. From the validation, it is found that GuardRider reduce the BER and FER by at least an order of magnitude. The throughput is also presented in Fig.~\ref{fig:fer} by taken the frame error rate into account. As expected, GuardRider also achieves significant improvement in the performance of backscatter systems. Furthermore, GuardRider is excited by the real WiFi traffics generated using {\it ping} to send packet with size $2048$ bytes and interval $100\mu$s, of which the results is presented in Fig.~\ref{fig:fer}(c). It is found that GuardRider can still backscatter information, while the baseline system is unusable in this scenario.

\section{Discussion}
{\it The possibility of using WiFi transceiver to decode the message.}
In this work, we use an USRP as a receiver to decode the message to ease the implementation of our system. However, our system can be implemented on top of the proposed code translation technique to enable the decoding at the COTS WiFi transceivers. It should be emphasized here that the primary aim of GuardRider is to solve the low reliability caused by the silent period of the excitation signals. Therefore, we did not implement GuardRider receiver using WiFi transceivers, which is left as a future study.

{\it Higher layer design.}
GuardRider mainly focuses on the physical layer design of the backscatter communication which aims to solve the problem caused by the intermittent nature of the WiFi signals. However, the higher layer protocols should also match the properties of the physical layer protocols, such as how does the tag efficiently perform RS coding and how does the receiver inform the results of the parameter estimation to the tag. We plan to propose a protocol stack for GuardRider and make it online available.

\section{Related Work}
GuardRider is a WiFi backscatter system which firstly aims to improve the reliability of backscatter communication over WiFi signals in the wild. We use the the statistical knowledge about the alternation of on-and-off states of the WiFi signal in the wild and optimize an RS code to follow the transition of WiFi states. The WiFi backscatter system has been intensively studied recently in \cite{FreeRider, BackFi, Bryce14, Kellogg16, HitchHike, Kim18, Zhang16}.

{\bf WiFi backscatter.} BackFi \cite{BackFi} operates backscatter communication over the WiFi excitation signals transmitted from WiFi APs with hardware modification. Reference \cite{Bryce14} connected the RF-powerd devices to the Internet excited by a WiFi signal. Passive Wi-Fi demonstrates for the first time it is able to generate 802.11b backscatter transmissions using backscatter communications \cite{Kellogg16}. HitchHike \cite{HitchHike} and \cite{Zhang16} enable the backscatter communication over 802.11b signals of the COTS WiFi transceivers using the proposed codeword translation technique. FreeRider \cite{FreeRider} further extends the backscatter communication over other excited RF radios, such as Bluetooth, 802.11g/n WiFi and ZigBee. Recent work \cite{Kim18} enables per-symbol and in-band backscatter communication over the WiFi excitation signals using a so-called flicker detector by utilized the residual channel knowledge of the WiFi packets. Besides, backscattering the ultra-wideband signals is considered in \cite{Yang17}. However, the reliability is rarely considered in these systems. In order to enable reliable backscatter communication, the inherent nature of the excitation signal should be considered.   

{\bf RS code in backscatter.} EraRFID \cite{EraRFID} enables identification of missing tags by reading only a subset of the tag based on the RS codes. Reference \cite{Burmester16} evaluates the theoretical lower bound for the redundancy of the erasure code and the memory-erasure tradeoff for recovering missing tags using RS codes. The RS coded identification information of the tag can be stored into the RFID tag by the reader, which is different to our system. Our system adopts RS codes with adaptive coding rates to protect backscatter communication and we implemented the RS encoder with FPGA.  
\section{Conclusion}
WiFi backscatter communication suffers from low reliability due to the intermittent nature of the WiFi traffics. GuardRider is the first backscatter system which aims to improve the reliability of the backscatter communicaiton riding over WiFi signals. We proposed an optimization algorithm for designing RS code to overcome the low reliability problem, and implemented GuardRider using FPGA. It was found from the results that GuardRider significantly reduce the bit error rate and frame error rate. Hence, GuardRider is highly reliable even the excitation signal having the randomly alternated on and off states. We plan to implement GuardRider by using COTS WiFi routers and propose a protocol stack as a future study.

\ifCLASSOPTIONcaptionsoff
  \newpage
\fi

\bibliographystyle{IEEEtran}

\bibliography{reference}

\end{document}